\documentclass[prl,aps,superscriptaddress,footinbib,twocolumn,floatfix]{revtex4-2}
\usepackage[latin9]{inputenc}
\usepackage{color}
\usepackage{amsmath}
\usepackage{amssymb}
\usepackage{stmaryrd}
\usepackage{csquotes}
\usepackage{graphicx}
\usepackage[unicode=true,bookmarks=true,bookmarksnumbered=false,bookmarksopen=false,breaklinks=false,pdfborder={0 0 1},backref=false,colorlinks=true]
 {hyperref}
\hypersetup{
 linkcolor=magenta, urlcolor=blue, citecolor=blue, pdfstartview={FitH}, unicode=true}

\usepackage{siunitx}
\usepackage{multirow}
\usepackage{graphicx}

\usepackage{subfiles}
\usepackage{tikz}
\usetikzlibrary{quantikz2}
\usepackage{algorithm}
\usepackage{algorithmic} 
\makeatletter

\usepackage{amsfonts}\usepackage{tabularx}\usepackage{dcolumn}\usepackage{bm}\usepackage{graphicx}\usepackage{epstopdf}
\usepackage{times}
\hypersetup{urlcolor=blue}

\def\ket#1{\left|#1\right\rangle}

\def\Bc{{\beta_{\text{C}}}}
\def\Bq{{\beta_{\text{Q}}}}

\makeatother

\begin{document}
\title{Improved Nonlocality Certification via Bouncing between Bell Operators and Inequalities}

\affiliation{Center for Quantum Information, IIIS, Tsinghua University, Beijing 100084, China\\
$^2$ Instituut-Lorentz, Universiteit Leiden, P.O. Box 9506, 2300 RA Leiden, The Netherlands\\
$^{3}$ Zhejiang Key Laboratory of Micro-nano Quantum Chips and Quantum Control, School of Physics, Zhejiang University, Hangzhou 310027, China\\
$^{4}$ Shanghai Qi Zhi Institute, Shanghai 200232, China\\
$^{5}$ ZJU-Hangzhou Global Scientific and Technological Innovation Center, Hangzhou 310027, China\\
$^{6}$ Hefei National Laboratory, Hefei 230088, China}

\author{Weikang Li$^{1,2}$}\thanks{These authors contributed equally to this work.}
\author{Mengyao Hu$^{2}$}\thanks{These authors contributed equally to this work.}
\author{Ke Wang$^{3}$}
\author{Shibo Xu$^{3}$}
\author{Zhide Lu$^{4}$}
\author{Jiachen Chen$^{3}$}
\author{Yaozu Wu$^{3}$}
\author{Chuanyu Zhang$^{3}$}
\author{Feitong Jin$^{3}$}	
\author{Xuhao Zhu$^{3}$}
\author{Yu Gao$^{3}$}
\author{Zhengyi Cui$^{3}$}
\author{Aosai Zhang$^{3}$}	
\author{Ning Wang$^{3}$}
\author{Yiren Zou$^{3}$}
\author{Fanhao Shen$^{3}$}
\author{Jiarun Zhong$^{3}$}
\author{Zehang Bao$^{3}$}
\author{Zitian Zhu$^{3}$}
\author{Pengfei Zhang$^5$}
\author{Hekang Li$^{5}$}
\author{Qiujiang Guo$^{5}$}
\author{Zhen Wang$^{3,6}$}
\author{Dong-Ling Deng$^{1,4,6}$}
\author{Chao Song$^{3,6}$}
\email{chaosong@zju.edu.cn}
\author{H. Wang$^{3,5,6}$}
\author{Patrick Emonts$^{2}$}
\email{emonts@lorentz.leidenuniv.nl}
\author{Jordi Tura$^{2}$}

\begin{abstract}
{
Bell nonlocality is an intrinsic feature of quantum mechanics, which can be certified via the violation of Bell inequalities.
It is therefore a fundamental question to certify Bell nonlocality from experimental data.
Here, we present an optimization scheme to improve nonlocality certification by exploring flexible mappings between Bell inequalities and Hamiltonians corresponding to the Bell operators. 
We show that several Hamiltonian models can be mapped to new inequalities with improved classical bounds than the original one, enabling a more robust detection of nonlocality.
From the other direction, we investigate the mapping from fixed Bell inequalities to Hamiltonians, aiming to maximize quantum violations while considering experimental imperfections.
As a practical demonstration, we apply this method to an XXZ-like honeycomb-lattice model utilizing over {$70$} superconducting qubits. 
The successful application of this technique, as well as combining the two directions to form an optimization loop, may open new avenues for developing more practical and noise-resilient nonlocality certification techniques and enable broader experimental explorations.
}
\end{abstract}

\maketitle

Nonlocality, characterized by correlations that cannot be explained within a local hidden variable framework, is among the most mysterious and fundamental features of the quantum world~\cite{Einstein1935Can}.
This property can be revealed via the violation of a Bell inequality~\cite{Bell1964Einstein,Clauser1969Proposed,Brunner2014Bell},
where measurements on entangled quantum systems yield correlations surpassing the classical reach.
To this end, early-stage experiments were carried out to reveal such nonlocal aspects of the physical world, primarily focused on optical platforms~\cite{Freedman1972Experimental,Aspect1981Experimental,Weihs1998Violation,Rowe2001Experimental}.
Over the recent years, several loopholes in detecting nonlocality, especially the detection loophole and the locality loophole, have been carefully considered~\cite{Larsson2014Loopholes,Aspect2015Closing}.
With notable advances in quantum techniques, loophole-free experiments have been reported on various platforms, including nitrogen-vacancy centers~\cite{Hensen2015Loopholefree}, photons~\cite{Giustina2015SignificantLoopholeFree,Shalm2015Strong,Li2018Test}, neutral atoms~\cite{Rosenfeld2017EventReady}, and more recently superconducting circuits~\cite{Storz2023Loopholefree}.

Apart from its fundamental importance, 
nonlocality is also a crucial resource for various applications.
On the one hand, nonlocality enables device-independent information processing tasks and plays an important role in, e.g. quantum cryptography~\cite{Acin2007DeviceIndependent,Xu2020Secure,Pirandola2020Advances}, 
randomness amplification~\cite{Colbeck2011Private,Colbeck2012Free,Gallego2013Full},
and self-testing~\cite{Supic2020Selftesting}.
On the other hand, nonlocality serves as a key resource for unconditional computational advantages without any assumptions of complexity hierarchies~\cite{Bravyi2018Quantum,LeGall2019AverageCase}.
Furthermore, with the advent of quantum machine learning~\cite{Biamonte2017Quantum,Dunjko2018Machine,Li2022Recent}, 
the resource for quantum learning advantages is under active exploration.
Along this direction, rigorous proofs have been proposed, indicating that nonlocality is vital to learning advantages for both generative models~\cite{Gao2022Enhancing} and classification models~\cite{Zhang2024QuantumClassical}.

In practical scenarios, especially many-body systems, experimental observations of Bell nonlocality face the challenge of noise~\cite{Tura2014Detecting,Schmied2016Bell,Engelsen2017Bell,Wang2017Entanglement,Tura2017Energy,Guo2023Detecting,Frerot2023Probing,Bernards2023Bell}.
By assigning measurement settings to different parties in a Bell inequality, 
we obtain a Bell operator. This operator can be equivalently treated as a Hamiltonian, where the states with energy lower than the classical bound would indicate nonlocality~\cite{Tura2017Energy}.
However, due to the presence of inevitable experimental noise, 
the prepared states may deviate from the ideal state, with their observed energies failing to cross the classical-quantum boundary.

\begin{figure*}[t]
\center
\includegraphics[width=0.9\linewidth]{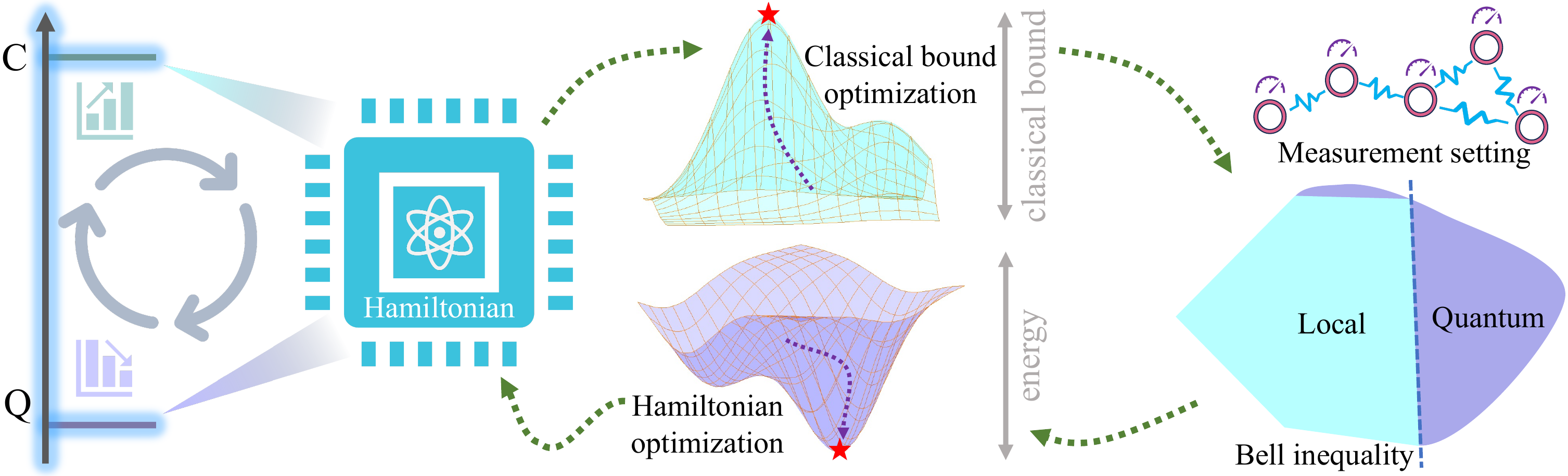}
\caption{A schematic illustration of the optimization scheme executed between Bell inequalities and Hamiltonians. 
We begin by parameterizing both Bell inequalities and measurement settings, and a Bell inequality where the quantum measurement observables replace the measurements forms an equivalent Hamiltonian $\mathcal{H}$. 
For a specified spin Hamiltonian,
by keeping $\mathcal{H}$ fixed and equal to this spin Hamiltonian,
we search through the parameter space to find Bell inequalities with higher classical bounds (upper half).
For a specified Bell inequality,
with experimentally measured correlators,
we can also optimize the measurement settings to find lower quantum values (lower half). 
This process forms a loop and can be executed iteratively to maximize the gap between the classical bound and the quantum value until it converges.
}
\label{fig1}
\end{figure*}

In this paper, we propose an optimization approach to address this problem by employing flexible mappings between Bell inequalities and Hamiltonians, where the latter corresponds to the so-called Bell operator.
For concreteness,
for a fixed Hamiltonian, there might be numerous corresponding Bell inequalities equipped with different quantum measurement settings.
We first parameterize the measurements.
The coefficients of the inequality and Hamiltonian can be connected through mappings determined by the measurement settings.
With gradient-based methods, we search the parameter space for inequalities with improved classical bounds.
Our results show that the selected Hamiltonians, connected to some Bell inequalities from previous works including a variant of Gisin's elegant inequality~\cite{Gisin2009Bell,Tura2017Energy}, can be mapped to new inequalities with notably more favorable classical bounds through optimization. 
This method also applies in certain cases where the
Hamiltonian corresponds to the optimal quantum measurements (Supplemental Material, App.~C~\cite{Note0}).
To demonstrate the application of this framework,
we experimentally implemented a variational quantum algorithm to search for low-energy states of an XXZ-like honeycomb-lattice model on a $73$-qubit superconducting processor. 
The classical bound of the original Bell inequality fails to be violated due to experimental noise, 
while a tailored inequality specifically designed can certify nonlocality with a high confidence level,  without requiring the use of extra data from the experiment.

From the other direction,
for a fixed Bell inequality and a set of pre-measured correlators,
we optimize the measurement settings to obtain improved quantum values~\cite{Liang2007Bounds,Poderini2022Initio,Xu2023Graphtheoretic}.
We demonstrate that the optimized settings exhibit larger violations and higher robustness under different strengths of depolarizing noise.
Combining these two directions,
we can optimize the Hamiltonian of a given Bell inequality to obtain lower quantum values, and then optimize the inequality corresponding to this optimized Hamiltonian for higher classical bounds, thus yielding a larger gap between the quantum and classical limits of the inequalities obtained.
This procedure can be executed iteratively as a loop,
where a schematic illustration of our framework is exhibited in Fig.~\ref{fig1}.

{\it Preliminaries.}---The basic Bell inequality model we consider here is a variant~\cite{Tura2017Energy} of Gisin's elegant Bell inequality~\cite{Gisin2009Bell}.
This setting describes a bipartite scenario,
where party $A$ has four dichotomic measurements with outcomes $\pm 1$ and $B$ has three with outcomes $\pm 1$.
Parameterized by $\Delta$, this inequality has the following classical bound~\cite{Note0}
\begin{equation}
\label{eq:Gisin_delta_Bell}
\begin{aligned}
   \mathcal{I}_G(\Delta) \geqslant -2|\Delta|-|\Delta+2|-|\Delta-2|,
\end{aligned}
\end{equation}
where it becomes $-8$ with $\Delta=2$.
We will use this setting for both numerical and experimental explorations.
By assigning quantum measurements to the different parties in this inequality (see Supplemental Material, App.~B~\cite{Note0} for details),
we equivalently obtain the two-body Hamiltonian
\begin{equation}
\label{eq:Gisin_delta_Ham}
\begin{aligned}
   \mathcal{H}_G =\frac{4}{\sqrt{3}}\left(\sigma_x^{(1)} \sigma_x^{(2)}+\sigma_y^{(1)}\sigma_y^{(2)}+2\sigma_z^{(1)}\sigma_z^{(2)}\right),
\end{aligned}
\end{equation}
where $\sigma_x^{(i)}$, $\sigma_y^{(i)}$ and $\sigma_z^{(i)}$ are Pauli operators acting on the qubit indexed by $i$~\cite{Note0}.
This Hamiltonian has the ground state energy about $-9.24$, exceeding the classical bound.
This model can be generalized to many-body systems, such as one-dimensional chains~\cite{Tura2017Energy} and two-dimensional lattices~\cite{Emonts2024Effects}, by applying appropriate couplings between different qubits.

{\it From Hamiltonians to Bell inequalities.}---We begin with the parameterization of quantum measurement settings of different parties.
A single-qubit measurement $\Vec{n}\cdot\Vec{\sigma}$, 
where $\Vec{n}$ denotes the measurement direction and $\Vec{\sigma}=\{\sigma_x,\sigma_y,\sigma_z\}$ denotes the Pauli basis, 
can be visualized in the Poincar\'{e} sphere shown in Fig.~\ref{fig2}.
For all measurements, we denote the parameters collectively as $\boldsymbol{\theta}$.
For a fixed Hamiltonian, we first decompose it into the Pauli strings and use $\Vec{h}$ to collectively denote its coefficients in the Pauli basis.
We also use a vector $\Vec{\alpha}$ to represent the Bell inequality, where $\Vec{\alpha}_{i,j}$ denotes the coefficient for the correlator between party $i$ and $j$.
In this formalism, the Bell inequality and the Hamiltonian can be connected through a matrix $\mathbf{T}(\boldsymbol{\theta})$
as a function of measurement parameters $\boldsymbol{\theta}$~\cite{Emonts2024Effects} (see Supplemental Material, App.~A):
\begin{equation}
\label{eq:HamToBell}
\begin{aligned}
   \mathbf{T}(\boldsymbol{\theta}) \cdot\Vec{\alpha}=\Vec{h}.
\end{aligned}
\end{equation}
This equation may be under-determined depending on the number of measurements in the Bell scenario. 
Upon obtaining a solution for the inequality determined by $\Vec{\alpha}$, the corresponding classical bound can be calculated.

\begin{figure*}[t]
\center
\includegraphics[width=1\linewidth]{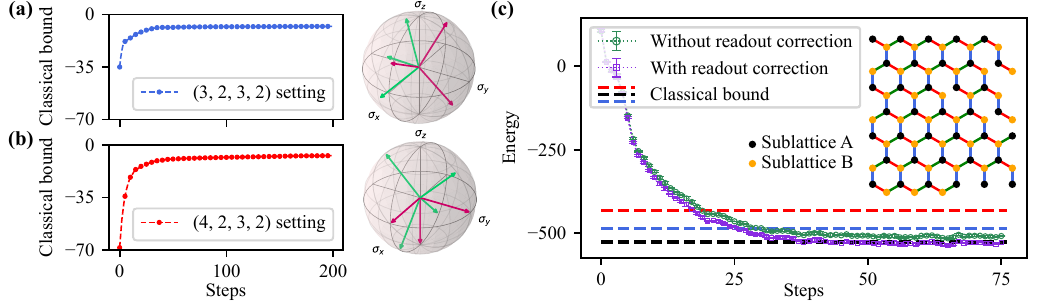}
\caption{Mappings from fixed spin Hamiltonians to Bell inequalities.
(a-b) Optimized classical bounds as a function of training steps, where the Poincar\'{e} sphere on the right side exhibits the measurement settings corresponding to the highest classical bound.
Cyan and pink arrows denote measurement observables of party $A$ and $B$, respectively.
(a) considers parties $A$ and $B$ with three dichotomic measurements with outputs $\pm1$;
(b) considers parties $A$ and $B$ with four and three dichotomic measurements with outputs $\pm1$, respectively.
(c) Experimental variational quantum optimization for the two-dimensional honeycomb model with $73$ superconducting qubits. We present the measured energy of the model Hamiltonian with respect to training steps and the classical bound for the original classical bound (black dashed line from Eq.~\eqref{eq:Gisin_delta_Bell}).
The optimized classical bound is further plotted in blue dashed line for (a) and red dashed line for (b), respectively.
}
\label{fig2}
\end{figure*}

To formalize this problem as an optimization task, we define the cost function as the classical bound to be maximized, i.e., $\mathcal{L} = \mathcal{C}(\boldsymbol{\theta})$, where $\mathcal{C}(\boldsymbol{\theta})$ represents the classical bound obtained from Eq.~\eqref{eq:HamToBell} for a given set of measurement parameters $\boldsymbol{\theta}$. 
By maximizing the cost function, we aim to find the optimal set of measurement parameters that yields the highest possible classical bound. This optimization problem can be handled by gradient-based optimization algorithms.
Here, we apply gradient ascent with the Adam optimizer.
The first task we work on is focused on a variant of Gisin's elegant inequality, with the classical bound mentioned in Eq.~\eqref{eq:Gisin_delta_Bell}.
Given the Hamiltonian in Eq.~\eqref{eq:Gisin_delta_Ham},
we apply the optimization procedure to find new Bell inequalities with higher classical bounds.

Let the $(m_a,d_a,m_b,d_b)$ Bell scenario denote that party $A$ and $B$ have $m_a$ and $m_b$ measurements with $d_a$ and $d_b$ outcomes, respectively.
We consider two scenarios.
In the first scenario, both parties $A$ and $B$ have three dichotomic measurements with outputs $\pm1$, i.e., the $(3,2,3,2)$ scenario.
The Adam optimization is executed for $10^4$ steps and the parameters corresponding to the highest classical bound are recorded as the final output.
We plot the first $200$ steps in Fig.~\ref{fig2}(a), where the training converges in dozens of steps.
For the selected parameters,
the classical bound achieved is around $-7.39$, notably higher than the original bound of $-8$.
To visualize the optimized measurements,
we also use the Poincar\'{e} sphere to exhibit the optimized measurement directions in Fig.~\ref{fig2}(a).
For the second $(4,2,3,2)$ scenario, party $A$ has four dichotomic measurements while $B$ has three.
By keeping the other hyperparameters the same, we execute the optimization and plot the results in Fig.~\ref{fig2}(b).
The achieved highest classical bound is around $-6.56$, further improving the value obtained in the first scenario by a notable ratio.

{\it Application to many-body experiments.}---We build a quantum spin model in a honeycomb grid as shown in Fig.~\ref{fig2}(c), which is implemented by $73$ superconducting qubits arranged on an $11\times11$ square lattice.
These selected qubits feature median fidelities of simultaneous single- and two-qubit gates around $99.95\%$ and $99.4\%$, respectively~\cite{Note0}.
The Hamiltonian model is a generalized version of Eq.~\eqref{eq:Gisin_delta_Ham} by assigning the two-body Hamiltonian at nearest-neighbor sites with coupling strengths $J_{\kappa}(\epsilon)$,
\begin{equation}
\label{eq:HamXXZ}
\begin{aligned}
   \mathcal{H}=\sum_{\kappa\text{-link}} \frac{4}{\sqrt{3}}J_{\kappa}(\epsilon)\left(\sigma_x^{(m)} \sigma_x^{(n)}+\sigma_y^{(m)}\sigma_y^{(n)}+2\sigma_z^{(m)}\sigma_z^{(n)}\right)_{\kappa}, \nonumber
\end{aligned}
\end{equation}
where the local observables act on the two linked qubits indexed $m$ and $n$.
For the red links in Fig.~\ref{fig2}(c), the coupling strength is $J_\kappa(\epsilon)=1+\epsilon$, while it is $(1-\epsilon)/2$ for green and blue links.
The Bell inequality is designed by summing over the edges of the honeycomb lattice, which has a similar structure as the Hamiltonian.
We set $\epsilon=0.94$ in our experiment.
Since all the coupling strengths are positive and translation-invariant, the optimal classical deterministic strategy can be achieved by local optimal strategies of these connected pairs.

To search for low-energy states of the honeycomb-lattice model, we apply variational quantum algorithms and set the target as preparing the ground states~\cite{Peruzzo2014Variational,McClean2016Theory,Kokail2019Selfverifying,Cerezo2021Variational,Bharti2022Noisy}.
The parameterized quantum circuit we utilized has the same structure as Ref.~\cite{Wang2024Probing}. 
We calculate the gradients directly through quantum measurements according to the parameter shift rule \cite{Mitarai2018Quantum,Ren2022Experimental}, and use the Adam optimizer to update the circuit parameters.
After $60$ iteration steps, the training curve is shown in Fig.~\ref{fig2}(c).
We find that for the original Bell inequality in Eq.~\eqref{eq:Gisin_delta_Bell},
the energy of experimentally prepared states
is around $-514$, which
cannot violate the classical bound.
Even with corrections for readout errors, the minimal value achieved is around $-533$, which is slightly lower than the classical bound but the confidence level due to statistical shot noise is low.
To handle this problem, 
we apply the previously found two Bell inequalities and extend them to the many-body system.
As exhibited in the blue line in Fig.~\ref{fig2}(c), 
by considering the Bell inequality from Fig.~\ref{fig2}(a),
the classical bound can be increased from $-526$ to $-486$, which allows the experimentally prepared states to violate the inequality without readout correction methods.
Furthermore, when switching to the inequality adapted from Fig.~\ref{fig2}(b),
the classical bound is further improved to $-431$.
This enables certifying nonlocal correlations with a notably higher ($3$x) confidence level.

{\it Optimization-assisted analytical results.}---In the above results, we have applied optimization-based methods to search for Bell inequalities corresponding to a fixed Hamiltonian.
Interestingly, in Fig.~\ref{fig2}(b), 
we find the numerical results possess good symmetry to provide analytical results.
For concreteness, 
as shown on the right side, the four measurement directions of party $A$ form a regular tetrahedron, while the three of party $B$ are mutually orthogonal.
The inferred optimized inequality (not tight~\cite{Note11}) is written as
\begin{equation}
\label{eq:analytical_Bell}
\begin{aligned}
   \mathcal{I}_{\text{opt}} &=A_0B_1 -A_0B_2 - A_1B_0 + A_1B_2 \\
   &+ A_2B_0 + A_2B_2 - A_3B_1 - A_3B_2
   \geqslant -4
\end{aligned}
\end{equation}
up to a constant factor, where the four measurements of party $A$ are $\frac{1}{\sqrt{2}}(\sigma_y+\sigma_z)$, $\frac{1}{\sqrt{2}}(-\sigma_x-\sigma_z)$, $\frac{1}{\sqrt{2}}(\sigma_x-\sigma_z)$, $\frac{1}{\sqrt{2}}(-\sigma_y+\sigma_z)$ and the three measurements of party $B$ are $\sigma_x$, $\sigma_y$, $\sigma_z$.
According to this formulation, this result can be regarded as playing two CHSH games for $\{A_0,A_3;B_1,B_2\}$ and $\{A_1,A_2;B_0,B_2\}$ together, and taking a summation.

{\it From Bell inequalities to Hamiltonians.}---We have studied optimizable mappings from Hamiltonians to Bell inequalities, which enables more robust experimental observations of nonlocality.
On the other side, it is also of practical importance to explore the mapping from Bell inequalities to Hamiltonians.
To certify nonlocality in experiments after preparing the entangled states,
measurements on a set of correlators constitute the observations.
However, due to the existence of experimental noise as discussed above, 
certifications may fail.
The idea of mapping Bell inequalities to Hamiltonians concerns the situation where we already have the measured values of these correlators.
According to Eq.~\eqref{eq:HamToBell}, for a fixed Bell inequality, it is possible to optimize the measurement settings to form a new Hamiltonian.
In this case, since we already have the experimental values of different correlators, we do not need to solve the ground state energy but replace these correlators with measured values.
Let a row vector $\Vec{c}$ represent the set of measured values of quantum correlators, the target becomes minimizing
$\Vec{c}\cdot\Vec{h}=\Vec{c}\cdot\mathbf{T}(\boldsymbol{\theta})\cdot\Vec{\alpha}$ over $\boldsymbol{\theta}$.

We choose the inequality in Eq.~\eqref{eq:Gisin_delta_Bell} and Hamiltonian in Eq.~\eqref{eq:Gisin_delta_Ham},
and carry out simulations of such mappings concerning depolarizing quantum noise, i.e., in each operation for preparing the target state, we attach a noise channel
\begin{equation}
\label{eq:noise}
\begin{aligned}
   \mathcal{N}_p(\rho)=(1-3p) \rho+p\left(
   \sigma_x\rho \sigma_x+\sigma_y\rho \sigma_y+\sigma_z\rho \sigma_z \right)
\end{aligned}
\end{equation}
for the state $\rho$, where $p$ denotes the noise strength.
As shown in Fig.~\ref{fig3}(a), we optimize the Hamiltonian according to states prepared through channels with varying noise strengths (see Supplemental Material, App.~B~\cite{Note0}).
The measured energy for the original Hamiltonian is increasingly affected by higher depolarizing noise $p$, and the violation disappears at $p=0.008$.
Meanwhile, the optimized Hamiltonian showcases higher robustness:
For each noise strength, although the Bell inequality and classical bound are fixed,
the quantum value is lower and the violation disappears only at $p=0.012$.

\begin{figure}[t]
\center
\includegraphics[width=1\linewidth]{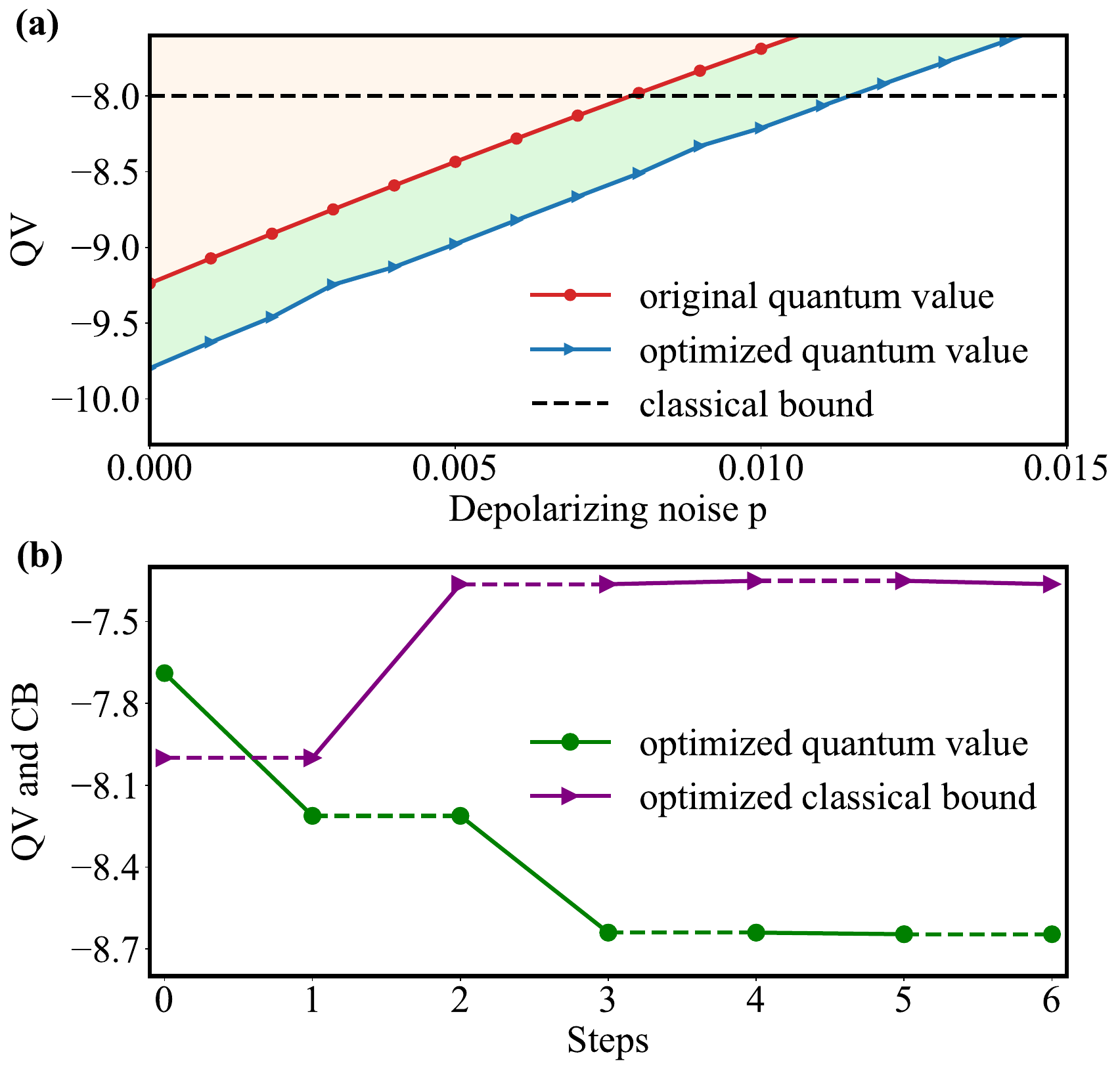}
\caption{(a) Original quantum values (QV) and optimized quantum values considering different depolarizing noise strengths, where the black dashed line denotes the classical bound.
(b) Iteratively optimizing both the Bell inequality and the Hamiltonian to maximize the gap between the classical bound (CB) and quantum value.
The dashed purple line denotes that the Bell inequality remains fixed at this step while optimizing the measurement settings;
Similarly, the dashed green line denotes that the Hamiltonian is fixed at this step during which the inequality is optimized.
The optimization converges in two loops of iterations.
}
\label{fig3}
\end{figure}

With optimizable mappings for both directions established,
we can iteratively apply the optimization to both sides.
The starting point we consider is the above model with depolarizing noise $p=0.010$, where the original inequality fails to be violated due to the effect of noise as shown in Fig.~\ref{fig3}(a).
The procedure begins with optimizing the measurement settings to compile a new Hamiltonian with lower energy.
Then for this Hamiltonian,
we optimize the Bell inequality for higher classical bounds while keeping the Hamiltonian fixed.
These two steps form a loop which we repeatedly execute several times until converging to a stable point.
The iterative optimization process is exhibited in Fig.~\ref{fig3}(b).
After two loops,
the classical bound and quantum value converge, and the larger gap between them enables better certification of nonlocality.

We remark that the experimental data needs to be obtained only once, assuming the data allows for the estimation of any two-body correlator; e.g. via randomized Pauli strings. 
Otherwise, we would have to supplement existing data with new measurements for the freshly chosen measurement settings.

{\it Discussion and outlook.}---The connection between Bell inequalities and Hamiltonians provides a useful tool for studying nonlocal correlations in quantum systems. 
In this work, we have presented a framework for optimizing mappings between Bell inequalities and Hamiltonians. 
This provides higher experimental feasibility and robustness to noise for detecting Bell nonlocality. 
By parameterizing the quantum measurements and applying gradient-based optimization methods, 
we have shown that it is possible to find measurement settings that lead to significantly higher classical bounds.
This enlarges the gap between the classical bound and quantum value, ultimately providing more amenable certification of nonlocality in experiments. 
We further carry out experimental demonstrations in a many-body honeycomb-lattice model on a superconducting processor, in which nonlocality could not be certified in the original scenario.
By utilizing the optimization methods, the experimental demonstrations reveal the existence of nonlocal correlations. 
The results demonstrate the effectiveness of this approach in practical scenarios, where experimental noise significantly affects the observed nonlocal behaviors. 
Furthermore, these optimization-assisted analytical results also contribute to the theoretical understanding of nonlocality in quantum systems and may provide guidance for future experimental designs.

On the other hand,
by assigning appropriate measurement settings to the Bell inequality, we can derive a corresponding Hamiltonian whose ground state energy serves as a witness for certifying nonlocality. 
By exploring the parameter space of the measurement settings for a fixed inequality, we can identify optimal measurement configurations that help build a Hamiltonian with minimal quantum values.
The combination of the above two directions provides a loop-optimization scheme, which further improves the performance.

The optimization framework presented in this work opens up several avenues for future research. 
A challenge that is encountered towards optimizing more complex scenarios is the accumulation of local minima in the $\boldsymbol{\theta}$ parameter space. 
This could possibly be mitigated by applying cutting-edge machine-learning techniques from deep reinforcement learning~\cite{Fawzi2022Discovering} or large language models.
For Bell inequalities or quantum systems featuring certain symmetries,
group equivariant machine learning models may be utilized to explore the solution in the symmetry-restricted space~\cite{Cohen2016Group}.
Last, in this work we apply classical optimization techniques for nonlocality problems.
With the advances in variational quantum algorithms~\cite{Cerezo2021Variational},
it is also appealing to develop hybrid quantum-classical optimization algorithms to probe the interplay between Bell inequalities and quantum systems.

While we have addressed the optimization of many-body systems by focusing only on relatively small-size subsystems, it would be highly desirable to apply this framework to increasingly larger subsystems and investigate the scaling behaviors. 
This includes correlators involving more parties and a larger number of measurements for an individual party.
We have exemplified the usefulness of our framework in turning experimental data that is quantum, but could not be certified so, into certifiably non-local correlations, simply as a post-processing step that does not require extra experimental work.

\vspace{.2cm}
\noindent\textbf{Note:} All the data and code for this study will be organized and made publicly available for download on Zenodo/Figshare/Github upon publication.

\vspace{.2cm}
\noindent\textbf{Acknowledgement:} We thank  Vedran Dunjko, Marc-Olivier Renou, Benjamin Schiffer, Pei-Xin Shen, Wenjie Jiang, and Jin-Fu Chen for helpful discussions. 
P.E. and J.T. acknowledge the support received by the Dutch National Growth Fund
(NGF), as part of the Quantum Delta NL programme. 
P.E. acknowledges the support received through the NWO-Quantum Technology
programme (Grant No.~NGF.1623.23.006).
J.T. acknowledges the support received from the European Union's Horizon Europe research and innovation programme through the ERC StG FINE-TEA-SQUAD (Grant No.~101040729). 
This publication is part of the `Quantum Inspire - the Dutch Quantum Computer in the Cloud' project (with project number [NWA.1292.19.194]) of the NWA research program `Research on Routes by Consortia (ORC)', which is funded by the Netherlands Organization for Scientific Research (NWO).
{
The device was fabricated at the Micro-Nano Fabrication Center of Zhejiang University.  
We acknowledge support from the Innovation Program for Quantum Science and Technology (Grant Nos.~2021ZD0300200 and 2021ZD0302203), the National Natural Science Foundation of China (Grants Nos.~12174342, 12274367, 12322414, 12274368, U20A2076, 12075128, and T2225008), the National Key R\&D Program of China (Grant No.~2023YFB4502600), and the Zhejiang Provincial Natural Science Foundation of China (Grant Nos.~LDQ23A040001, LR24A040002)}. W.L., Z.L., and D.-L.D. are supported in addition by Tsinghua University Dushi Program, and the Shanghai Qi Zhi Institute.

The views and opinions expressed here are solely
those of the authors and do not necessarily reflect those of the funding institutions. Neither
of the funding institutions can be held responsible for them.

\footnotetext[0]{See Supplemental Material}

\footnotetext[11]{This inequality is the combination of two tight Bell inequalities. It is therefore not tight for the $(4,2,3,2)$ Bell scenario without considering the single-body terms $A_{i}$ and $B_{j}$. Moreover, the corresponding face of this inequality has a dimension of $10$, while the facets defined by the tight inequalities have a dimension of $11$.}

\bibliography{QMLBib}

\clearpage
\newpage 
\onecolumngrid
\setcounter{section}{0}
\setcounter{equation}{0}
\setcounter{figure}{0}
\setcounter{table}{0}
\setcounter{page}{1}
\makeatletter
\renewcommand\thefigure{S\arabic{figure}}
\renewcommand\thetable{S\arabic{table}}
\renewcommand\theequation{S\arabic{equation}}

\begin{center} 
	{\Large \bf Appendix}
\end{center} 

\setcounter{figure}{0}
\setcounter{table}{0}
\renewcommand\thefigure{S\arabic{figure}}
\renewcommand\thetable{S\arabic{table}}
\maketitle

\textbf{\textit{Roadmap:}}
Appendix~\ref{supp:nonlocality} provides detailed background on utilizing the energy as a witness to detect nonlocality and reviews relevant works,
followed by an introduction to the general framework of the optimization protocol in the main text.
The detailed hyperparameter settings are provided in Appendix~\ref{supp:numsetting}.
In Appendix~\ref{supp:numextend}, we provide additional numerical simulations.
Appendix~\ref{supp:device} provides information about the superconducting processor utilized to demonstrate variational quantum state preparations on a $73$-qubit two-dimensional honeycomb lattice.

\tableofcontents
\subsection{Energy as a witness of nonlocality and optimizable mappings}
\label{supp:nonlocality}

The connection between Bell inequalities and Hamiltonians provides an avenue to test nonlocality through certain engineered quantum systems~\cite{Tura2017Energy,Tura2014Detecting,Baccari2017Efficient,Wang2017Entanglement,Tavakoli2022Bell,Renou2019Genuine}. 
For a given Bell inequality, 
we can construct a Hamiltonian by replacing the measurements in the inequality with certain measurement settings.
If the ground state energy of this Hamiltonian falls below the classical bound $\Bc$,
the observed energy can be a nonlocality witness~\cite{Tura2017Energy}.
Conversely, 
for a specified spin Hamiltonian,
there might be numerous corresponding Bell inequalities with certain measurement settings. 
In this work, we aim to find desirable Bell inequalities and measurement settings, such that the gap between $\Bq$ and $\Bc$ is maximal by searching the coefficient space of Bell inequalities and parameters of measurement settings.
We explain the optimization scheme from two directions in the following two subsections.

\subsubsection{Bell inequality \texorpdfstring{$\rightarrow$}{->} Hamiltonian}
Given a Bell inequality $\mathcal{I}-\Bc \geqslant 0$, we first construct a Hamiltonian that coincides with its Bell operator. 
Here we consider the $(m_1,2,m_2,2)$ scenario (two parties, party $A$ ($B$) can perform $m_1$ ($m_2$) measurements and each measurement produces $2$ outcomes). 
The general Bell operator of the $(m_1,2,m_2,2)$ scenario can be written as
\begin{align}
\mathcal{B}_{\text{gen}}=\sum_{x_1,x_2=0}^{m_1-1,m_2-1}\sum_{k_1,k_2=0}^{1}\alpha_{x_1,x_2}^{(k_1,k_2)} A_{x_1}^{(k_1)}B_{x_2}^{(k_2)},
\end{align} 
where $A_{x_1}^{(k_1)}=\sum_{a=0}^{1}(-1)^{a k_1}F_{x_1, a}$ is the discrete Fourier transform of a positive operator-valued measure (POVM) $\{F_{x_1,a}\}_{a=0}^{1}$ representing the measurement on the party $A$ in the basis $x_1$ and  $B_{x_2}^{(k_2)}$ for party $B$ is defined similarly. 
Note that $[A_{x_1}^{(k_1)},B_{x_2}^{(k_2)}] = 0$ for $x_1 \in [m_1]=\{0,1,\cdots,m_1-1\}$, $x_2\in [m_2]=\{0,1,\dots,m_2-1\}$, and $k_1,k_2=0,1$ since they act on different systems.
Moreover, the term $A_{x_1}^{(k_1)}B_{x_2}^{(k_2)}$ simplifies to the local term $B_{x_2}^{(k_2)}$ when $k_1=0$. 
In this paper, we do not consider the local terms, so we can omit the superscripts $(k_1)$ and $(k_2)$ from $A_{x_1}^{(k_1)}B_{x_2}^{(k_2)}$. 
By fixing the operator, one can build a corresponding Hamiltonian $\mathcal{H}_{\text{gen}}$ such that $\mathcal{H}_{\text{gen}}\equiv \mathcal{B}_{\text{gen}}$.

As an example, consider the CHSH inequality, $\mathcal{I}=A_0B_0+A_0B_1+A_1B_0-A_1B_1$.
By letting $A_0=B_0=\sigma_x$, $A_1=B_1=\sigma_y$, then the Bell operator is
\begin{align}
\mathcal{B}_{\text{CHSH}}&=\sigma_x^{(1)}\sigma_x^{(2)}+\sigma_x^{(1)}\sigma_y^{(2)}+\sigma_y^{(1)}\sigma_x^{(2)}-\sigma_y^{(1)}\sigma_y^{(2)} \\
&=2\left(\frac{1}{2}(\sigma_x+\sigma_y)^{(1)}(\sigma_x+\sigma_y)^{(2)}-\sigma_y^{(1)}\sigma_y^{(2)}\right)
\end{align}
We can write $\mathcal{B}_{\text{CHSH}}$ in a way similar to $XY$ Hamiltonian by letting $\sigma_{\pi/4}=\frac{\sqrt{2}}{2}(\sigma_x+\sigma_y)=\cos \frac{\pi}{4} \sigma_x+\sin\frac{\pi}{4} \sigma_y$.
In this way, we obtain the Hamiltonian in the form of~\cite{Tura2017Energy}
\begin{align}
    \mathcal{H}_{\text{CHSH}} = \mathcal{B}_{\text{CHSH}}=2\left(\sigma_{\pi/4}^{(1)}\sigma_{\pi/4}^{(2)}-\sigma_y^{(1)}\sigma_y^{(2)}\right).
\end{align}

\subsubsection{Hamiltonian \texorpdfstring{$\rightarrow$}{->} Bell inequality}

In this section, we study the map from a Hamiltonian to Bell inequalities, and use the model adapted in the main text as an illustrative example.
We are given a Hamiltonian with two-body interactions at nearest-neighbor sites with coupling strengths $J_\kappa (\epsilon)$,
\begin{equation}
\label{eq:HamXXZ2}
\begin{aligned}
   \mathcal{H}(\Delta)=\sum_{\kappa\text{-link}} \frac{4}{\sqrt{3}}J_\kappa (\epsilon)\left(\sigma_x^{(m)} \sigma_x^{(n)}+\sigma_y^{(m)}\sigma_y^{(n)}+\Delta\sigma_z^{(m)}\sigma_z^{(n)}\right)_{\kappa},
\end{aligned}
\end{equation}
where $\Delta$ is a real parameter and
$\kappa$-link denotes the links between two connected qubits index $m$ and $n$.

Our goal is to find a Bell inequality and a measurement setting,
such that the
Bell operator $\mathcal{B}$ corresponds to this given Hamiltonian:
\begin{equation} \label{matrix_equation}
    \mathcal{B} \equiv \mathcal{H}(\Delta).
\end{equation}
If we restrict to the local part $\mathcal{H}_{\text{local}}= \frac{4}{\sqrt{3}}(\sigma_x^{(m)} \sigma_x^{(n)}+\sigma_y^{(m)}\sigma_y^{(n)}+\Delta\sigma_z^{(m)}\sigma_z^{(n)})$ of the Hamiltonian  $\mathcal{H}(\Delta)$, the structure of $\mathcal{H}_{\text{local}}$ requires a specific Bell scenario: the number of parties in local parts is two because of the tensor form of $\mathcal{H}_{\text{local}}$, and the number of outcomes $d =2$ due to the local dimension of the Pauli matrices. Thus for the local part $\mathcal{H}_{\text{local}}$, the scenario is $(m_1,2,m_2,2)$. In order to have non-trivial correlations, we set $m_i \geqslant 2,i=1,2$. 

According to the general form of Bell expression in the $(m_1,2,m_2,2)$ scenario, the Bell operator $\mathcal{B}_{\text{local}}$ without local terms associated 
 to $\mathcal{H}_{\text{local}}$ can be written as 
\begin{align}
\label{eq:general}
\mathcal{B}_{\text{local}}=\sum_{x_1,x_2=0}^{m_1-1,m_2-1}\alpha_{x_1,x_2}A_{x_1}B_{x_2}.
\end{align} 
Note that $\mathcal{B}=\sum_{\kappa\text{-link}}J_\kappa (\epsilon) (\mathcal{B}_{\text{local}})_{\kappa}$.

In general, a single-qubit operator can be written as 
\begin{equation}
    A_{x_1} = \vec{n}\cdot\vec{\sigma} = n_x\sigma_x + n_y\sigma_y + n_z\sigma_z,
\end{equation}
where the coefficient vector $\vec{n}=(n_x,n_y,n_z)$ is normalized: $n_x^2 + n_y^2 + n_z^2 = 1$. Thus the parameterized measurements of party $A$ and $B$ can be written as:
\begin{equation}
\label{eq:AB}
\begin{aligned}
A_{x_1} & =\cos \theta_{x_1} \sigma_x+\sin \theta_{x_1} \cos \phi_{x_1} \sigma_y +\sin \theta_{x_1} \sin \phi_{x_1} \sigma_z\\
B_{x_2} & =\cos \gamma_{x_2} \sigma_x+\sin \gamma_{x_2} \cos \varphi_{x_2} \sigma_y +\sin \gamma_{x_2} \sin \varphi_{x_2} \sigma_z,
\end{aligned}
\end{equation}
where $\theta_{x_1},\gamma_{x_2} \in [0,\pi]$, $\phi_{x_1} ,\varphi_{x_2} \in [0,2\pi)$ are variational real parameters and $x_1\in [m_1],x_2\in [m_2]$. 
In this case, by inserting Eq.~\eqref{eq:AB} to the general form of Bell inequality of the $(m_1,2,m_2,2)$ scenario as Eq.~\eqref{eq:general}, the Bell operator $\mathcal{B}_{\text{local}}$ is 
\begin{equation}
\label{eq:belloperator}
    \begin{split}
        \mathcal{B}_{\text{local}}  =& \sum_{x_1,x_2=0}^{m_1-1,m_2-1}\alpha_{x_1,x_2}A_{x_1}B_{x_2}\\
=&\sum_{x_1,x_2=0}^{m_1-1,m_2-1}\alpha_{x_1,x_2}\left(\cos \theta_{x_1} \sigma_x+\sin \theta_{x_1} \cos \phi_{x_1} \sigma_y +\sin \theta_{x_1} \sin \phi_{x_1} \sigma_z\right)  \\
&\otimes \left(\cos \gamma_{x_2} \sigma_x+\sin \gamma_{x_2} \cos \varphi_{x_2} \sigma_y +\sin \gamma_{x_2} \sin \varphi_{x_2} \sigma_z\right),
    \end{split}
\end{equation}
where $\alpha_{x_1,x_2}$ are the coefficients of correlators $A_{x_1}B_{x_2}$.
To find the coefficients $\alpha_{x_1,x_2}$ of the Bell expression $\mathcal{I}$ of the Bell inequality $\mathcal{I}-\Bc \geqslant 0$, we write the above Bell operator as a system of linear equations by projecting into the basis $\{\sigma_i\otimes \sigma_j \mid i,j\in \{x,y,z\}\}$. 
The projection of $\mathcal{H}_{\text{local}}$ is given by $ \operatorname{Tr}((\sigma_i\otimes \sigma_j)\mathcal{H}_{\text{local}})$, $i,j \in \{x,y,z\}$.
In this way, we obtain a system of linear equations as follows:
    \begin{align}
    \label{eq:belleq}
         \mathbf{T}(\boldsymbol{\theta})\cdot \vec{\alpha}=\vec{h},
    \end{align}
where 
\begin{equation}
   \mathbf{T}(\boldsymbol{\theta})=\begin{pmatrix}
       \cos \theta_0 \cos \gamma_{0} & \dots & \cos \theta_{x_1} \cos \gamma_{x_2} & \dots &\cos \theta_{m_1-1} \cos \gamma_{m_2-1} \\
       \cos \theta_0 \sin \gamma_{0}\cos\varphi_{0}& \dots& \cos \theta_{x_1} \sin \gamma_{x_2}\cos\varphi_{x_2}& \dots &\cos \theta_{m_1-1} \sin \gamma_{m_2-1}\cos\varphi_{m_2-1}  \\
       \vdots & \ddots & \ddots & \ddots & \vdots \\ 
       \sin\theta_{0} \sin \phi_{0} \sin \gamma_{0}\sin\varphi_{0} &\dots &\sin\theta_{x_1} \sin \phi_{x_1} \sin \gamma_{x_2}\sin\varphi_{x_2}&\dots   & \sin\theta_{m_1-1} \sin \phi_{m_1-1} \sin \gamma_{m_2-1}\sin\varphi_{m_2-1}
  \end{pmatrix}
\end{equation} is a $9\times m_1m_2$ matrix,
\begin{equation}
\vec{\alpha}=(\alpha_{0,0},\alpha_{0,1},\dots ,\alpha_{x_1,x_2},\dots ,\alpha_{m_1-1,m_2-1})^T, \notag
\end{equation}
and
\begin{align}
    \vec{h}=&( \operatorname{Tr}((\sigma_x\otimes \sigma_x)\mathcal{H}_{\text{local}}), \dots, \operatorname{Tr}((\sigma_i\otimes \sigma_j)\mathcal{H}_{\text{local}}), \dots, \operatorname{Tr}((\sigma_z\otimes \sigma_z)\mathcal{H}_{\text{local}}))^T, \quad i,j\in \{x,y,z\}. \notag
\end{align}
The matrix  $\mathbf{T}(\boldsymbol{\theta})$ is fully determined by choosing $\boldsymbol{\theta}=(\theta_{0},\dots,\theta_{m_1-1},\phi_{0},\dots,\phi_{m_1-1},\gamma_{0},\dots,\gamma_{m_2-1},\varphi_{0},\dots,\varphi_{m_2-1})$.
Since $\operatorname{rank}(\mathbf{T}(\boldsymbol{\theta}))\leq \min\{ 9,m_1m_2 \}$, 
we have the following two cases:
\begin{enumerate}
    \item If $\mathbf{T}(\boldsymbol{\theta})$ is invertible ($m_1=m_2=3$ and $\operatorname{rank}(\mathbf{T}(\boldsymbol{\theta}))=9$), there is a unique solution for $\vec{\alpha}=\mathbf{T}(\boldsymbol{\theta})^{-1}\vec{h}$. 
In this case, the solution $\vec{\alpha}$ is unique, so the Bell expression corresponds to $\mathcal{H}_{\text{local}}$ with operators given by Eq.~\eqref{eq:belloperator} is also unique. 
\item If $\mathbf{T}(\boldsymbol{\theta})$ is not invertible, there exists a family of solutions for $\vec{\alpha}$. 
It means there are multiple Bell expressions corresponding to $\mathcal{H}_{\text{local}}$ with operators given by Eq.~\eqref{eq:belloperator}.
\end{enumerate}

\subsection{Details for the numerics}
\label{supp:numsetting}
Both optimizations, the minimization of the quantum value and the maximization of the classical bound by adapting the measurements, are non-convex.
Thus, their convergence depends on the choice of parameters in the numerical algorithms.
Here, we give detailed settings of the hyperparameters used in the numerical simulations.

\subsubsection{A variant of Gisin's elegant Bell inequality}

We provide the details of a variant of Gisin's elegant inequality introduced in the main text.
In this setting, $A$ has four dichotomic measurements and $B$ has three, both with outcomes $\pm 1$.
The inequality is expressed as
\begin{equation}
\label{eq:gisin_var}
\begin{aligned}
   \mathcal{I}_G(\Delta)&=A_0 B_0 + A_1 B_0  -  A_2 B_0  -  A_3 B_0 + A_0 B_1  -  A_1 B_1  +  A_2 B_1  -  A_3 B_1 +\Delta A_0 B_2  - \Delta A_1 B_2  - \Delta A_2 B_2  + \Delta A_3 B_2 \\
   & \geqslant -2|\Delta|-|\Delta+2|-|\Delta-2|.
\end{aligned}
\end{equation}
By assigning four measurements
\begin{equation}
\label{eq:gisin_a}
\begin{aligned}
A_0=\frac{\sigma_x+\sigma_y+\sigma_z}{\sqrt{3}},  \quad
A_1=\frac{\sigma_x-\sigma_y-\sigma_z}{\sqrt{3}}, \quad
A_2=\frac{-\sigma_x+\sigma_y-\sigma_z}{\sqrt{3}},  \quad
A_3=\frac{-\sigma_x-\sigma_y+\sigma_z}{\sqrt{3}}, 
\end{aligned}
\end{equation}
to party $A$ and
three measurements
\begin{equation}
\label{eq:gisin_b}
\begin{aligned}
B_0&=\sigma_x, \quad
B_1=\sigma_y,  \quad
B_2=\sigma_z,
\end{aligned}
\end{equation}
to party $B$,
We obtain the Bell operator in the main text when $\Delta=2$.

\subsubsection{Hamiltonian \texorpdfstring{$\rightarrow$}{->} Bell inequality}

To map a fixed XXZ-like Hamiltonian to Bell inequalities (cf., Fig.~2 in the main text),
we start with a random initialization of measurement settings $\boldsymbol{\theta}$ to provide the initial $\mathbf{T}(\boldsymbol{\theta})$.
For the case $(m_1,2,m_2,2)=(3,2,3,2)$,
$\mathbf{T}(\boldsymbol{\theta})$ is invertible and $\vec{\alpha}=\mathbf{T}(\boldsymbol{\theta})^{-1}\vec{h}$ from Eq.~\eqref{eq:belleq} has a unique solution.
This solution determines the Bell inequality, and further determines the classical bound.
Thus, we use $\mathcal{C}(\mathbf{T}(\boldsymbol{\theta}))$ to represent the resulting classical bound. 
The classical bound can be alternatively computed using the tensor network contraction via tropical algebra~\cite{Hu2022Tropical,Emonts2024Effects,Hu2024Characterizing}.

For the optimization, we apply gradient-based methods to search for $\boldsymbol{\theta}^*$ with the highest classical bound.
With finite difference methods, 
we calculate the gradient
$\nabla_{\bm{\theta}}\mathcal{C}(\mathbf{T}(\boldsymbol{\theta}))$. To illustrate the basic idea, we can use direct gradient ascent to maximize the classical bound $\mathcal{C}(\mathbf{T}(\boldsymbol{\theta}))$:
\begin{equation}
\bm{\theta}_{n+1}=\bm{\theta}_{n}+\epsilon\nabla_{\bm{\theta}}\mathcal{C}(\bm{\theta}_{n}),
\end{equation}
where $\epsilon$ is the learning rate and $\bm{\theta}_i$ denotes the parameters at training step $i$.
For better training performance,
in our work, we instead feed the gradients into the Adam($0.9$,$0.999$) optimizer with a learning rate of $0.02$ to optimize the parameters during this procedure.
Here, $0.9$ and $0.999$ denote the hyper-parameters determining the initial decay rates for the first- and second-moment estimates, respectively.

\begin{figure*}[t]
\center
\includegraphics[width=0.6\linewidth]{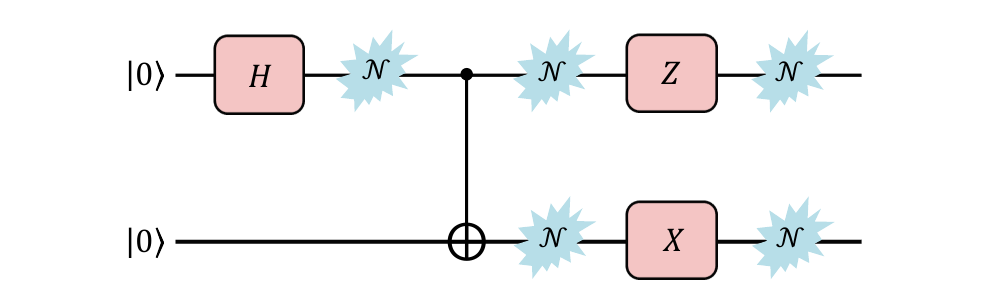}
\caption{The quantum circuit preparing the singlet state with the depolarizing noise introduced.
We use $\mathcal{N}$ to represent the depolarizing noise $\mathcal{N}_p(\rho)=(1-3p) \rho+p\left(
   \sigma_x\rho \sigma_x+\sigma_y\rho \sigma_y+\sigma_z\rho \sigma_z\right)$ for a given noise strength $p$.
}
\label{figs1}
\end{figure*}

For the case $(m_1,2,m_2,2)=(4,2,3,2)$,
Eq.~\eqref{eq:belleq} is underdetermined and there is a solution space.
Let $\mathcal{N}(\mathbf{T}(\boldsymbol{\theta}))$ denote the null space of $\mathbf{T}(\boldsymbol{\theta})$.
Provided any solution $\vec{\alpha}_0$ satisfying
Eq.~\eqref{eq:belleq} and any vector in the null space $\vec{u} \in \mathcal{N}(\mathbf{T}(\boldsymbol{\theta}))$,
\begin{equation}
\vec{\alpha} = \vec{\alpha}_0 + c \cdot \vec{u}
\end{equation}
is also a solution of Eq.~\eqref{eq:belleq}.
In this scenario,
in addition to optimizing the inequality and measurement setting, 
there is another dimension, i.e., the solution, to be optimized in the solution space.
In our optimizations, for the underdetermined equation's case, the solution with minimal norm suffices to provide good classical bound values.
We provide source code exploring maximal classical bounds in the whole solution space. 
Yet in practice, we use the minimal-norm solution for higher computational efficiency.

\subsubsection{Bell inequality \texorpdfstring{$\rightarrow$}{->} Hamiltonian}

For a given Bell inequality, we can optimize the measurement settings to search for lower quantum values.
In the numerical explorations of mapping Bell inequalities to Hamiltonians, 
the basic setting has been introduced in the main text.
Here, we provide more detailed information about noisy quantum simulations and the optimization scheme.

We first carry out the preparation of target states considering depolarizing noise.
For the model Hamiltonian [Eq.(1) in the main text],
the ground state is the singlet state $\frac{1}{\sqrt{2}}(\ket{01}-\ket{10})$.
We apply the circuit in {Fig.~\ref{figs1}} to prepare this state and add noise during the procedure.
By measuring the output states, we obtain the expectation values of all the two-body Pauli correlators and collect them as the simulated data.

In the main text, we use a row vector $\Vec{c}$ to represent the set of measured values of quantum correlators.
The target is to minimize the energy
\begin{equation}
\mathcal{E}(\boldsymbol{\theta})=\Vec{c}\cdot\mathbf{T}(\boldsymbol{\theta})\cdot\Vec{\alpha}.
\end{equation}
We can similarly apply gradient descent to search for measurement configurations to minimize it.
In practice, we feed the obtained gradients into the Adam($0.9$,$0.999$) optimizer with a learning rate of $0.01$.

\subsection{Extended numerics}
\label{supp:numextend}

\begin{figure*}[t]
\center
\includegraphics[width=0.8\linewidth]{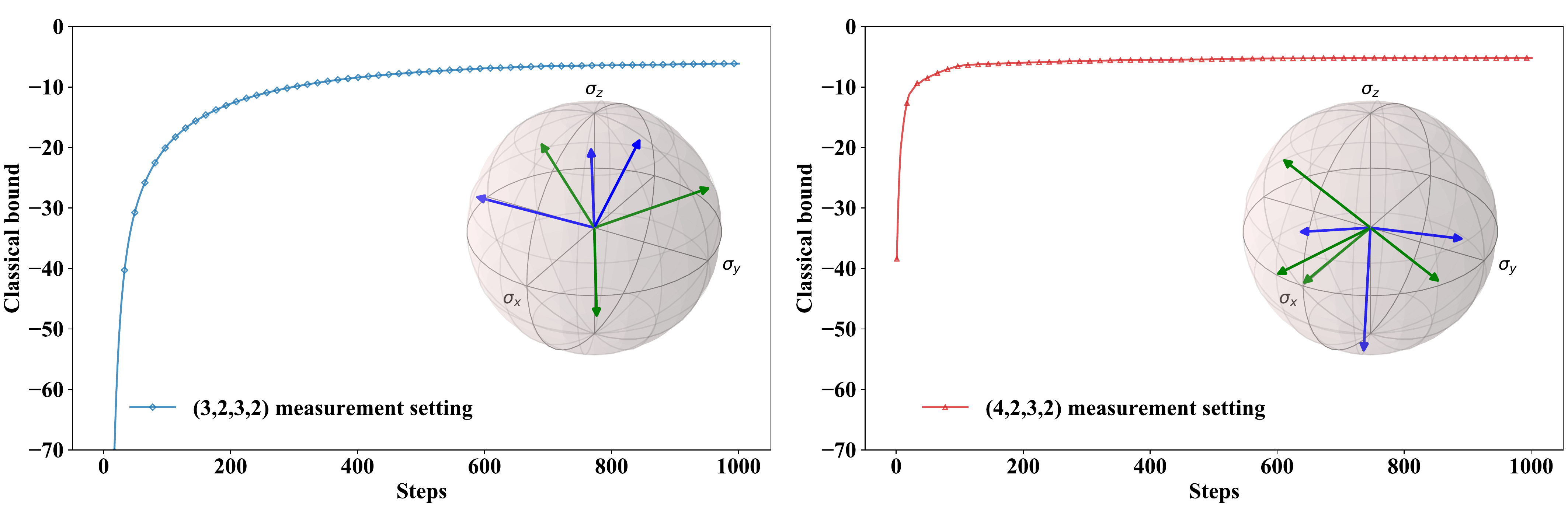}
\caption{Mappings from Hamiltonians to Bell inequalities.
We present optimized classical bounds as a function of training steps, as well as Poincar\'{e} spheres denoting the measurement settings corresponding to the highest classical bound.
Green and blue arrows denote measurement observables of party $A$ and $B$, respectively.
The left panel considers parties $A$ and $B$ with three dichotomic measurements with outputs $\pm1$, i.e., the $(3,2,3,2)$ scenario;
the right panel considers parties $A$ and $B$ with four and three dichotomic measurements with outputs $\pm1$, respectively, i.e., the $(4,2,3,2)$ scenario.
}
\label{figs2}
\end{figure*}

Here, we provide extended numerical experiments for mapping Hamiltonians to Bell inequalities.
We consider Gisin's elegant inequality~\cite{Gisin2009Bell}
\begin{equation}
\label{eq:gisin}
\begin{aligned}
   \mathcal{I}_{G0}&=A_0 B_0 + A_1 B_0  -  A_2 B_0  -  A_3 B_0 + A_0 B_1  -  A_1 B_1  +  A_2 B_1  -  A_3 B_1 + A_0 B_2  -  A_1 B_2  -  A_2 B_2  +  A_3 B_2 \\
   & \geqslant -6,
\end{aligned}
\end{equation}
which can be obtained by setting the $\Delta$ in Eq.~\eqref{eq:gisin_var} to be $1$.
By assigning four measurements of Eq.~\eqref{eq:gisin_a} to party $A$ and
three measurements of Eq.~\eqref{eq:gisin_b}
to party $B$,
we obtain the Bell operator
\begin{align}
    \label{eq:Gisin_op}
    \mathcal{B}_{G0}=\frac{4}{\sqrt{3}}\left(\sigma_x^{(1)} \sigma_x^{(2)}+\sigma_y^{(1)}\sigma_y^{(2)}+\sigma_z^{(1)}\sigma_z^{(2)}\right).
\end{align}
These quantum measurement observables are optimal for this inequality, i.e., they lead to the Bell operator with the minimal ground state energy.
We start from the fixed operator in Eq.~\eqref{eq:Gisin_op} and search for inequalities with higher classical bounds.
The results are exhibited in Fig.~\ref{figs2}.
For the $(m_1,2,m_2,2)=(3,2,3,2)$ scenario,
the classical bound is improved from $-6$ to $-5.58$.
For the $(m_1,2,m_2,2)=(4,2,3,2)$ scenario,
the classical bound is improved from $-6$ to $-5.17$.

\begin{table}[]
\caption{Performance of the $73$ qubits. The single-qubit gate error and CZ gate error are calibrated through cross-entropy benchmarking (XEB). 
Note that the gate errors are average errors rather than Pauli errors. The readout error is the average readout error for the $0$ state and the $1$ state.}
\begin{tabular}{l|l}
\hline
Parameter & Median Value \\ \hline
 T$_1$ & \SI{81.4}{\micro\second}   \\
 T$_2$ spin echo & \SI{23.5}{\micro\second}  \\
 Single-qubit gate error &  $4.54\times10^{-4}$ \\
 CZ gate error & $6.34\times10^{-3}$  \\
 Readout error & $8.72\times10^{-3}$  \\
 \hline
\end{tabular}
\label{tb:property}
\end{table}

\subsection{Device information}
\label{supp:device}

The superconducting quantum processor used in this work consists of $121$ qubits arranged in an $11$ by $11$ square lattice upgraded from the device employed in Ref.~\cite{Xu2023nonAbelian,Xu2024NonAbelian,Wang2024Probing}.
This processor is capable of performing a universal quantum gate set formed by \SI{30}{\nano\second} arbitrary single-qubit gates and \SI{40}{\nano\second} two-qubit CZ gates. 
We chose and benchmarked $73$ qubits to carry out the experiment, their properties are summarized in Table~\ref{tb:property}.
The optimization scheme follows the parameter-shift strategy from Ref.~\cite{Mitarai2018Quantum,Ren2022Experimental,Wang2024Probing}.

\end{document}